\documentclass[
 reprint,
 amsmath,amssymb,
 aps,
]{revtex4-2}

\usepackage{graphicx}
\usepackage{dcolumn}
\usepackage{bm}
\usepackage[mathscr]{euscript}
\usepackage{mathtools}
\usepackage{physics}
\usepackage{upgreek}
\usepackage[dvipsnames]{xcolor}
\usepackage{soul}
\usepackage{microtype}
\usepackage[hidelinks]{hyperref}

\newcommand{\dkpar}{\Delta{k}^{\parallel}}

\newcommand{\GUp}{\Gamma_{\uparrow}}
\newcommand{\GDown}{\Gamma_{\downarrow}}

\newcommand{\ai}{\alpha}
\newcommand{\ei}{\varepsilon}

\newcommand{\GDownMi}{\varepsilon_{i}}
\newcommand{\GUpMi}{\alpha_{i}}

\newcommand{\GDownMOne}{\varepsilon_{1}}
\newcommand{\GUpMOne}{\alpha_{1}}

\newcommand{\GDownMTwo}{\varepsilon_{2}}
\newcommand{\GUpMTwo}{\alpha_{2}}

\newcommand{\GTotDown}{\Gamma^{\mathrm{tot}}_{\downarrow}}
\newcommand{\GTotUp}{\Gamma^{\mathrm{tot}}_{\uparrow}}

\newcommand{\GThOne}{\Gamma_{\uparrow}^{\text{th}1}}
\newcommand{\GThTwo}{\Gamma_{\uparrow}^{\text{th}2}}

\newcommand{\GThTwoTilde}{\widetilde{\Gamma}_{\uparrow}^{\text{th}2}}

\definecolor{darkgreen}{RGB}{0,156,5}

\begin{document}

\preprint{APS/123-QED}

\title{Photonic Bose--Einstein condensation in the continuum limit}

\author{Andris Erglis}
 \thanks{andris.erglis@physik.uni-freiburg.de}
 \affiliation{Physikalisches Institut, Albert-Ludwigs-Universit\"at Freiburg, Hermann-Herder-Stra{\ss}e 3, 79104 Freiburg, Germany}

\author{Milan Radonji\'{c}}
\affiliation{Institut f\"ur Theoretische Physik, Universit\"at 
Hamburg, Notkestr.~9, 22607 Hamburg, Germany}
\affiliation{Institute of Physics Belgrade, University of Belgrade, 
Pregrevica 118, 11080 Belgrade, Serbia}

\author{Stefan Yoshi Buhmann}
\affiliation{
  Institut f\"ur Physik, Universit\"at Kassel, Heinrich-Plett Stra{\ss}e 40, 34132 Kassel, Germany
}

\date{\today}

\begin{abstract}
We investigate the properties of the photon Bose--Einstein condensate in the limit of small mode spacing. Alongside the well-known threshold of the phase transition at large mode spacings, we find an emergence of a second threshold for sufficiently small mode spacings, defining the crossover to a fully condensed state. Furthermore, we present our findings for the mode occupations in the precondensate or supercooling region towards the continuum limit.
\end{abstract}

\maketitle

The Bose--Einstein condensate (BEC) is a state characterized by a phase transition at a critical threshold above which the particles macroscopically occupy the lowest energy state. Typically, in experiments, the threshold for atomic BECs is the critical temperature \cite{pethick2008bose, davis1995bose}; for photons, it is the critical number of particles \cite{klaers2010bose, schmitt2018dynamics}.

In photon BECs, the setup normally consists of slightly curved mirrors forming a cavity with dye molecules in between the mirrors. A fixed cavity length provides a quasi two-dimensional (2D) setup with a transversal wavevector as the remaining degree of freedom. While a photon BEC can be achieved in such a confined setup, it is well known that in two dimensions, no condensation occurs in the thermodynamic limit, as the critical temperature (particle number) approaches zero (infinity) \cite{may1959superconductivity, hohenberg1967existence, bagnato1991bose, Kirsten1996}.

While many experimental and theoretical works study photon BECs within harmonic confinement mentioned above \cite{klaers2010bose, schmitt2014observation, marelic2015experimental, schmitt2015thermalization, schmitt2016spontaneous, damm2017first, radonjic2018interplay, ozturk2019fluctuation, ozturk2021observation, ozturk2023fluctuation}, recently, the photon BEC has also been demonstrated in a finite planar system with a 2D box potential \cite{busley2022compressibility}. Theoretically, an exhaustive description of the thermodynamical properties of a BEC has been given in \cite{Li2015}, also demonstrating the formation of a BEC in a 2D box potential.

The difference between small and large system sizes has been previously considered in a 2D harmonic oscillator setup \cite{nyman2018bose, Rodrigues2021}. It was demonstrated via the photon-BEC rate equations that for smaller mode spacings, the ground state occupation has a sharper increase at the threshold, exhibiting a sign of thermodynamic limit. Another limit of highly anisotropic three-dimensional geometries has been studied with massive bosons, where the phase transition into different condensation regimes was observed \cite{beau2010quasi}.

For a single mode, an analytical treatment of the PBEC rate equations in adiabatic approximation has been given \cite{bennett2020symmetry}. For small cavity decay rates, the photon number was found to be zero below a critical threshold while exhibiting a sharp transition into a condensate regime above the threshold with linear growth of the photon number.

The study of condensation in the continuum limit is also relevant to other systems with small mode spacing, e.g., chiral environments \cite{bennett2020symmetry}. In such cases, the condensate properties will qualitatively differ compared to larger mode spacings. Without appropriate treatment of small spacings, one initially would not be able to see the formation of the condensate. Instead, the condensate will only emerge for experimental parameters far from the ones expected for large mode spacings.

In this letter, we report our findings on the steady-state behavior of the photon BEC in a planar cavity at the crossover towards the continuum limit. Extending the analytical treatment of the photon BEC rate equations in Ref.~\cite{bennett2020symmetry} to two modes, we find two critical thresholds marking the condensate formation. We corroborate our findings with numerical simulations and evaluate the critical number from thermodynamic principles for an infinite system. Finally, we demonstrate the behavior of photon occupation number in the precondensate or supercooling regime, which emerges in the limit of small mode spacing.

Our model consists of a planar cavity where the control parameter is the transversal mode spacing $\dkpar$ with the dispersion relation for the cavity mode frequency $\omega (j_x, j_y, \dkpar)=c\epsilon^{-1/2}\sqrt{\left(j_z\pi/L_z\right)^2+(j_x^2+j_y^2)(\dkpar)^2}$, where $c$ is the speed of light, $\epsilon$ is the relative permittivity of the cavity medium, $j_z$ is the longitudinal mode number, $L_z$ is the cavity length, $j_x$ and $j_y$ are transversal mode numbers in $x$ and $y$ directions, respectively, and $\dkpar$ is defined as $\dkpar=\Delta k_x = \Delta k_y = \pi/L$, where $L$ is the length of the cavity mirrors in $x$ and $y$ directions. The ground mode is characterized by values $j_x=j_y=1$. 

As an inducer for effective photon--photon interaction, dye molecules are commonly used. Here, we rely on Rhodamine 6G dye which fulfills the Kennard--Stepanov relation $\GUpMi=\GDownMi e^{\beta \hbar (\omega_i-\omega_{\text{ZPL}})}$ where $\alpha_i$ and $\varepsilon_i$ are the respective stimulated absorption and emission coefficients for mode $i$, $\omega_\text{ZPL}$ is the zero phonon line, and $\beta=1/k_B T$ is the inverse temperature with $T=300$ K. The Rhodamine 6G spectrum \cite{julian2024private}, together with the position of the cavity modes, is shown in the Supplemental Material \cite{supplementalMaterial}.
\begin{figure*}
    \centering
    \includegraphics[width=0.95\textwidth]{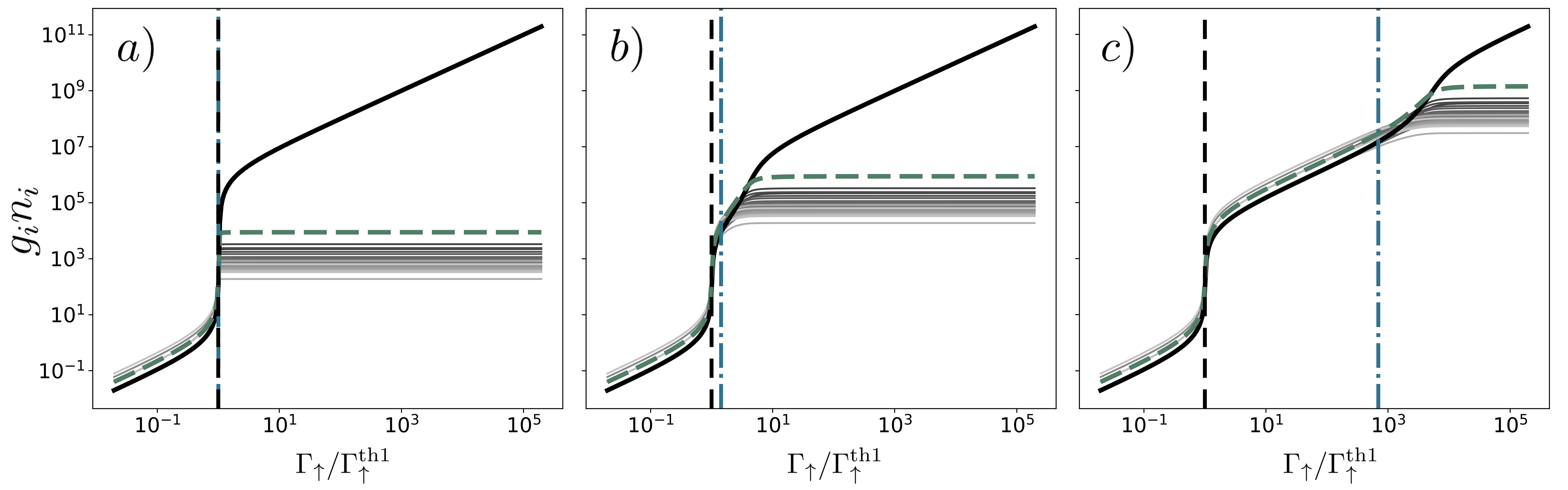}
    \caption{Mode populations in a steady state as a function of normalized pump rate where graphs from a) to c) are in order of decreasing mode spacing, $\dkpar=2\cdot10^4, 2\cdot 10^3, 50$ $\text{m}^{-1}$, respectively. The thick black and the green dashed curves are the ground and first excited states, respectively. The grey lines are all the other excited modes. The corresponding vertical dashed and dot--dashed lines show the value of the thresholds $\GThOne$ and $\GThTwo$. Here, $\GThOne\approx 5.1\cdot 10^6$ Hz. The simulations include 30 modes.}
    \label{fig:Plot1}
\end{figure*}  

As the longitudinal mode spacing between the $j_z$ and $j_z+1, j_z-1$ modes is comparable to the spectral width of the dye molecules, the cavity photons form a quasi two-dimensional system with fixed longitudinal mode number $j_z$.

The rate equations describing the dynamics of the PBEC can be obtained from the master equation using an open quantum systems description \cite{kirton2013nonequilibrium, buhmann2022nested}. In adiabatic approximation with respect to the state of the dye molecules, the rate equations for the photon numbers $n_i$ in mode $i$ are given by
\begin{equation} \label{eq:rate_equation}
\dot{n}_i(\GUp)=-\kappa n_i +M \frac{\GDownMi \GTotUp (n_i+1)-\GUpMi \GTotDown n_i}{\GTotUp+\GTotDown}.
\end{equation}
Here, $\kappa$ is the cavity decay rate, dependent on the reflectivity of cavity mirrors, $M$ is the total number of molecules, $\GTotUp=\GUp+\sum_{i=1}^{\mathscr{M}} g_i \GUpMi n_i$ is the total absorption rate with $\GUp$ being the pumping rate of the molecules and $g_i$ being the degeneracy of mode $i$. Similarly, $\GTotDown=\GDown+\sum_{i=1}^{\mathscr{M}} g_i \GDownMi (n_i+1)$ is the total emission rate where $\GDown$ is the the spontaneous decay rate of the molecules. Photons emitted via spontaneous decay are lost from the system into non-cavity modes. The $n_i+1$ terms are responsible for stimulated and spontaneous decay into cavity mode $i$.

For an analytical treatment, we restrict ourselves to the smallest number of modes needed to observe the second threshold, $\mathscr{M}=2$. To find the pump threshold frequency at which the condensation forms, we first algebraically solve the steady-state condition $\dot{n}_2=0$ for $n_2=n_2(n_1)$. The explicit form is written in \cite{supplementalMaterial}. Next, we solve $\big(\partial^3 n_2 /\partial \GUp^3\big )_{n_1}=0$ to obtain $\GUp=\GUp(n_1)$. This approach is inspired by the fact that the second derivative of the single mode solution $\dd^2 n_1 /\dd \GUp^2$ in the limit of small cavity decay \cite{bennett2020symmetry} gives a Dirac delta function at the first threshold position $\GUp=\GThOne$ \cite{supplementalMaterial}. The explicit solution \cite{supplementalMaterial} marks the special points where $n_2$ changes its behavior and indicates the condensation point. To proceed further, we solve the single-mode equation $\dot{n}_1=0$ \cite{bennett2020symmetry} and find $n_1(\GUp)$ \cite{supplementalMaterial}. We substitute the solution into the threshold condition for $\GUp=\GUp(n_1)$ and solve for $\GUp$, leading to two solutions as threshold conditions. Expanding them to the first order in $\kappa/M$ which is a small parameter compared to $\ai_i, \,\ei_i$, and assuming  $\GDownMTwo\approx\GDownMOne$, and $\GUpMTwo\approx\GUpMOne$ for small mode spacings, they read:
\begin{equation}\label{eq:threshold}
\begin{aligned}
   &\GThOne = \frac{(\GDown+\GDownMOne)\GUpMOne}{\GDownMOne},\\
   &\GThTwo = \frac{\GDown \GUpMOne}{\GDownMOne}+\frac{\GUpMOne (\GDownMOne+\GUpMOne)  \kappa }{M \abs{\GDownMTwo \GUpMOne-\GDownMOne  \GUpMTwo}}.
\end{aligned}
\end{equation}

The first threshold $\GThOne$ is similar as in Ref.~\cite{bennett2020symmetry}, except for the extra $\GDownMOne$ term in the brackets. The second threshold $\GThTwo$ is separated from $\GThOne$ by an amount linearly dependent on $\kappa$, inversely proportional to $M$, and inversely proportional to the difference between absorption/emission rates. This implies that at infinitesimal mode spacing, the denominator tends to zero, and consequently, $\GThTwo$ approaches infinity. It is noteworthy to mention that each threshold is mainly governed by one of two characteristic loss channels in our system: The first threshold is determined by the spontaneous emission of molecules $\GDown$, while the second threshold is induced by the cavity decay $\kappa$.

We now turn to numerical simulations of the photon population for more modes and demonstrate the steady state behavior for different mode spacings together with the thresholds from Eq.~\eqref{eq:threshold}.  
\begin{table}[b]
\caption{\label{tab:table1}
Parameters used in numerical simulations
}
\begin{ruledtabular}
\begin{tabular}{cll}
\textrm{Symbol}&
\textrm{Name}&
\textrm{Value}
\\
\colrule
$\kappa$ & Cavity decay &5 GHz\\
$\GDown$ & Spontaneous decay  & 244 MHz \\
$M$ & Number of molecules & $10^9$\\
$L_z$ & Cavity length & 1.489 $\upmu$m\\
$\sqrt{\epsilon}$ & Refractive index & 1.34\\
$j_z$ & Longitudinal mode number & 7\\
\end{tabular}
\end{ruledtabular}
\end{table}
Table \ref{tab:table1} shows the parameters we have used in this paper, which are comparable to the values used or measured in experiments. The steady-state photon populations as a function of pumping rate $\GUp$ are shown in Figure \ref{fig:Plot1} a) to c). In Fig.~\ref{fig:Plot1} a), the mode spacing is sufficiently large so that the first and second thresholds coincide. We observe a sharp increase in all populations at the first threshold, and for increasing pumping rate ground mode separates from all excited modes which saturate at constant values. In Fig.~\ref{fig:Plot1} b), we observe a separation of $\GThTwo$ from $\GThOne$. Furthermore, the behavior of populations is qualitatively different compared to Fig.~\ref{fig:Plot1} a): The excited mode populations saturate for higher pump values, while in ground mode populations we observe a kink above the second threshold. We refer to this region between both thresholds as the precondensate region where the condensate is not fully formed as the ground mode occupation grows in tandem with the occupation of the neighboring modes. Alternatively, this region may be referred to as the supercooling region in analogy with the supercooling of water between the freezing and the crystal homogeneous nucleation points. Finally, in Fig.~\ref{fig:Plot1} c), the second threshold is completely separated from the first threshold. All populations increase simultaneously, and above $\GThTwo$ only the ground mode continues to grow. We see that $\GThTwo$ marks the separation of the ground mode from the other modes for all three cases, while $\GThOne$ remains fixed. As we decrease $\dkpar$, a larger pump rate is required for full condensation. Ultimately, in the limit of the free 2D photon gas, the ground mode will only separate at an infinite pump rate.

\begin{figure}
    \centering
    \includegraphics[scale=0.51]{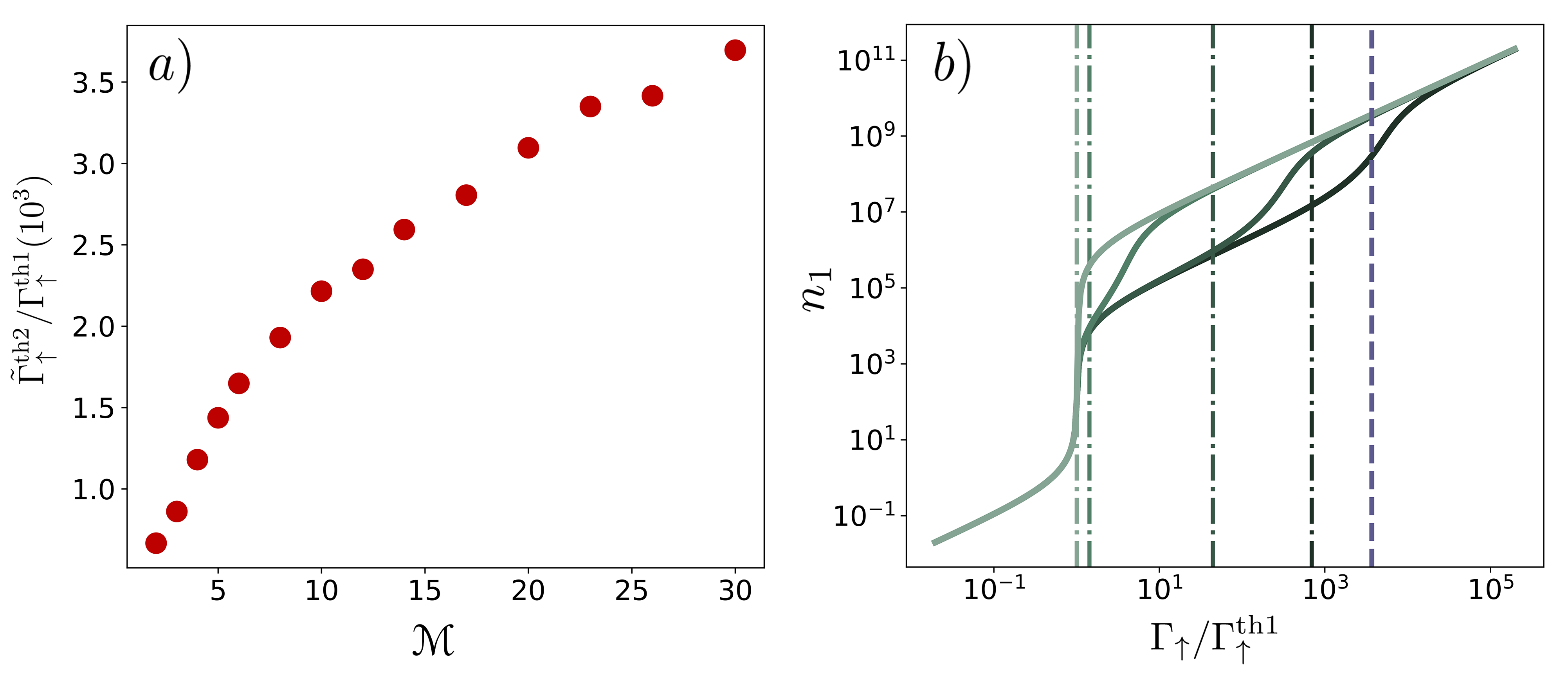}
    \caption{a) Value of the exact threshold $\GThTwoTilde$ as a function of the mode number $\mathscr{M}$. Here, $\dkpar=50$ $\text{m}^{-1}$. b) Ground mode populations simulated for 30 modes. Each curve depicts four different mode spacings in a decreasing value from left to right ($\dkpar=2\cdot 10^4, 2\cdot 10^3, 2\cdot 10^2, 50$ $\text{m}^{-1}$) (lighter to darker shades of green). Vertical dot--dashed lines indicate the second threshold $\GThTwo$ for each respective mode spacing. The vertical purple dashed line depicts $\GThTwoTilde$ for $\dkpar=50$ $\text{m}^{-1}$.}
    \label{fig:Plot2}
\end{figure}

The method of finding the zero-crossing of $\partial^3 n_2 /\partial \GUp^3$ allows us to numerically obtain the second threshold values for larger mode numbers, whose analytical expression is not attainable. We call it the exact second threshold $\GThTwoTilde=\GUp|_{\dd^3 n_2 /\dd \GUp^3=0}$ which marks an exact position of the condensation for any mode number. For $\mathscr{M}=2$, we have $\GThTwo\approx\GThTwoTilde$ and in Supplemental Material \cite{supplementalMaterial} we show a deviation of $\GThTwo$ from $\GThTwoTilde$.
Figure \ref{fig:Plot2} a) shows $\GThTwoTilde$ as a function of the mode number with a monotonic increase.
Figure \ref{fig:Plot2} a) shows that $\GThTwo$ marks the condensation point too early for larger mode numbers. Despite that, $\GThTwo$ can still qualitatively describe the threshold point for decreasing $\dkpar$, as shown in Fig.~\ref{fig:Plot2} b).
We see how $\GThTwo$ consistently marks the kink of the ground mode towards its linear condensed asymptote even for 30 modes. For comparison, we show the position of $\GThTwoTilde$ obtained for 30 modes. We see that $\GThTwoTilde$ marks the condensation point more precisely, as expected.

To analyze $\GThTwo$ further, we relate it to the critical particle number obtained by counting the photons in the Bose--Einstein distribution. We can analytically evaluate the critical photon number assuming a paraxial approximation, $k_x, k_y \ll k_z$. Following along the lines of \cite{chaba1975bose, kleinert2009path}, we obtain the critical photon number $N_{\text{c}}$ in a 2D system for small mode spacing $\dkpar$:
\begin{equation}\label{eq:critical_analyt}
    N^{(\infty)}_{\text{c}}=\!\!\! \sideset{}{'}\sum_{j_x, j_y=1}^{\infty} \! [e^{\beta \hbar [\omega(j_x, j_y)-\omega(1, 1)]}-1]^{-1}\approx - \frac{\pi}{2} \frac{1}{s} \ln{s},
\end{equation}
where $s=\beta \hbar c (\dkpar)^2/k_z \sqrt{\epsilon}$ and in the primed sum we have excluded the pair $(j_x, j_y)=(1,1)$. To relate the two thresholds from Eq.~\eqref{eq:threshold} with the photon number, we substitute the values into Eqs.~\eqref{eq:rate_equation} and solve for the total photon number in all the excited states $N_{\text{ex}}(\GUp)$, obtaining two critical photon numbers, $N_{\text{ex}}(\GThOne)\equiv N^{(1)}_{\text{c}}$ and $N_{\text{ex}}(\GThTwo)\equiv N^{(2)}_{\text{c}}$. For $\mathscr{M}=2$, we have $N_{\text{ex}}\equiv g_2 n_2$. 

\begin{figure}
    \centering
    \includegraphics[width=0.48\textwidth]{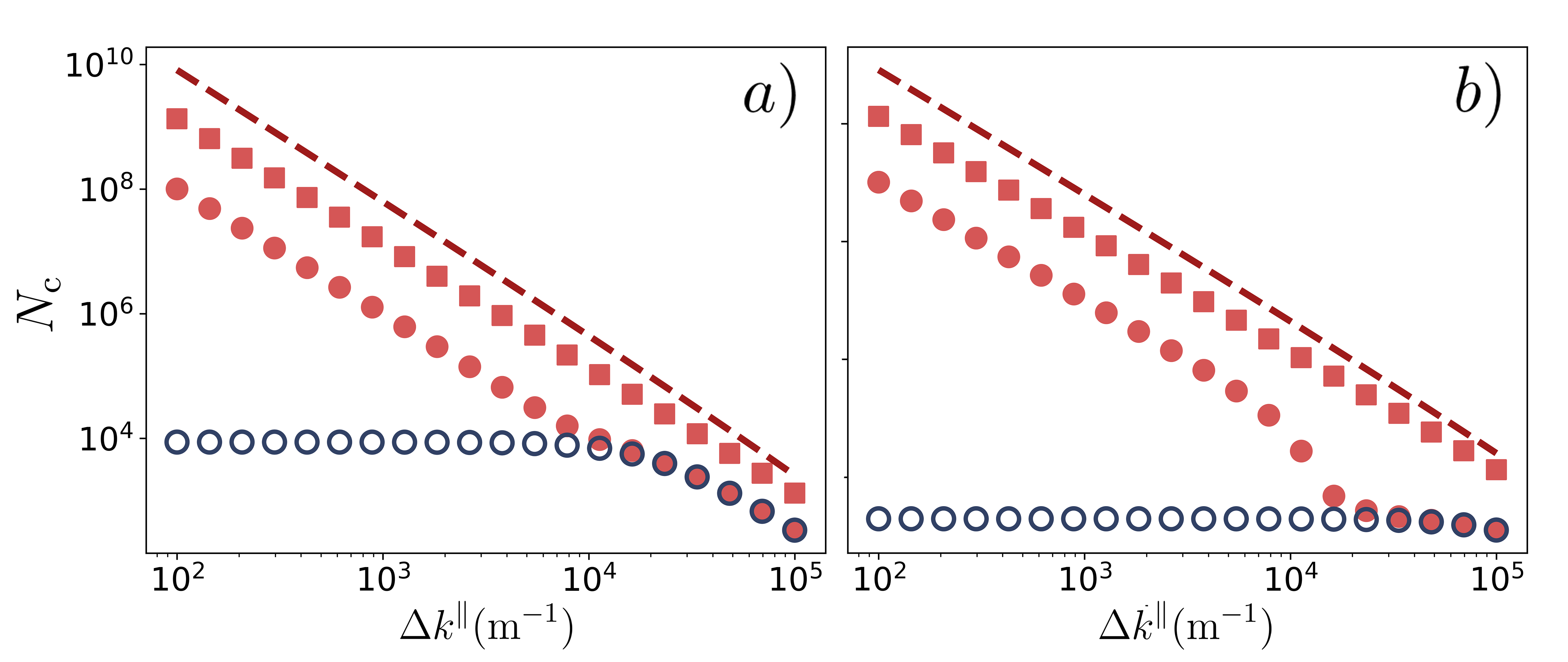}
    \caption{Critical number of particles as a function of the mode spacing $\dkpar$ with a) $\kappa=0.05$ GHz and b) $\kappa=5$ GHz. Red filled and blue hollow dots depict critical numbers for $\mathscr{M}=2$ obtained from values of $\GThTwo$ and $\GThOne$ in Eq.~\eqref{eq:threshold}, respectively. Red squares illustrate $N_c$ obtained from $\GThTwoTilde$ for $\mathscr{M}=30$.} The dashed line depicts thermodynamic solution from Eq.~\eqref{eq:critical_analyt}.
    \label{fig:Plot4}
\end{figure}

Figure \ref{fig:Plot4} compares the critical photon numbers $N^{(1)}_{\text{c}}$, $N^{(2)}_{\text{c}}$, and $N^{(\infty)}_{\text{c}}$ for decreasing mode spacing.
At small $\dkpar$ the behaviour of $N^{(2)}_{\text{c}}$ is qualitatively similar to $N^{(\infty)}_{\text{c}}$, i.e., both numbers tend to infinity as $\dkpar\to 0$. An offset is present between $N^{(\infty)}_{\text{c}}$ and $N^{(2)}_{\text{c}}$ which is expected since the critical photon number from Eq.~\eqref{eq:threshold} has been obtained by assuming only two modes in the cavity, while Eq.~\eqref{eq:critical_analyt} assumes infinitely many modes. Crucially, we see that $N^{(1)}_{\text{c}}$ fails to predict the critical number for small mode spacings as it saturates, similarly as in Fig.~\ref{fig:Plot1}. For large mode spacings, $N^{(1)}_{\text{c}}$ and $N^{(2)}_{\text{c}}$ coincide. Comparing Figs.~\ref{fig:Plot4} a) and \ref{fig:Plot4} b) we see that for smaller cavity decay rates, $N^{(1)}_{\text{c}}$ separates from $N^{(2)}_{\text{c}}$ at smaller $\dkpar$ value than for larger $\kappa$, consistent with Eq.~\eqref{eq:threshold}. Moreover, the qualitative behavior of $N^{(2)}_{\text{c}}$ around the separation point indicates higher nonlinearity for larger values of $\kappa$.

For comparison, in Fig.~\ref{fig:Plot4} we also show the critical photon number obtained with $\GThTwoTilde$ for 30 modes (also see Fig.~\ref{fig:Plot2}). Naturally, we observe that for higher mode numbers the critical photon number approaches $N^{(\infty)}_{\text{c}}$.

As seen in Fig.~\ref{fig:Plot1} and Fig.~\ref{fig:Plot2} b), for small $\dkpar$, the second threshold diverges and the supercooling region extends to larger pumping strengths. Therefore, it is insightful to investigate further the behavior of the difference between photons in the ground and excited modes. Since populations in the neighboring modes in the supercooling region are similar, as seen in Fig.~\ref{fig:Plot1} c), it is possible to utilize the perturbation method for a generic number of modes. We expand the occupation in mode $i$ perturbatively, $n_i= n_1+q_i\Delta n$ for $i\geq 2$, where $\Delta n$ is a small difference in photon number and $q_i$ are mode-dependent coefficients. We choose $q_2 = 1$ without loss of generality. In a similar manner, we express the absorption and emission rates as $\ai_i=\GUpMOne+\mu_i\Delta\Gamma$ and $\ei_i=\GDownMOne+\eta_i\Delta\Gamma$, where $\Delta\Gamma$ is a small deviation proportional to the mode spacing. Again, $\mu_i$ and $\eta_i$ are the mode-dependent coefficients where we choose $\mu_2\equiv 1$ and $\eta_2\equiv \eta$. With these assumptions and only retaining the lowest orders of $\Delta\Gamma$ and $\Delta n$, we can solve Eqs.~\eqref{eq:rate_equation} analytically for $\mathscr{M}=2$ or higher.
\begin{figure}
    \centering
    \includegraphics[scale=0.4]{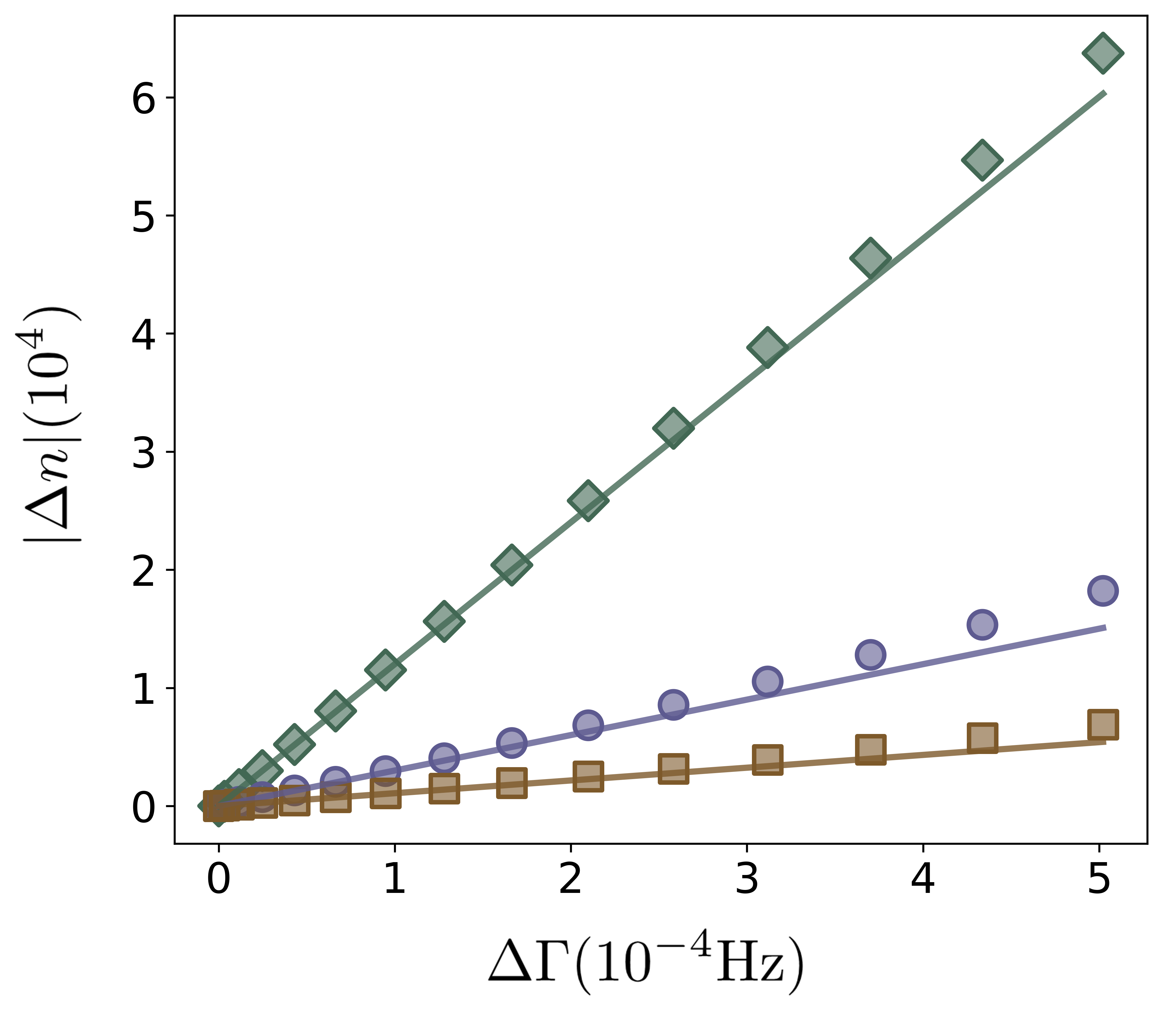}
    \caption{Absolute value of the difference between ground and first excited mode $\abs{\Delta n}$ as a function of $\Delta\Gamma$ for different mode numbers $\mathscr{M}=2$ (green rhomboids), $4$ (purple circles), $6$ (red squares). Solid lines are the analytical expression from Eq.~\eqref{eq:dn}. Here, the value of the pump rate is $\GUp=2\GThOne$.}
    \label{fig:Plot3}
\end{figure}
For a nondegenerate ground mode $g_1=1$, in the limit of small cavity decay $\kappa$ we obtain the solution for $\Delta n$:
\begin{equation}\label{eq:dn}
\Delta n=
   \begin{cases}
      \displaystyle n_1^2\frac{(\ei_1-\ai_1\eta)\Delta\Gamma}{\ai_1\ei_1}, & \GUp\geq\GThOne,\\
      0, & \text{otherwise}.
    \end{cases}   
\end{equation}
Here, $n_1$ is the photon occupation of the ground mode:
\begin{equation}\label{eq:n1_sol_dn_general}
n_1=
   \begin{cases}
      \displaystyle\frac{M(\ei_1 \GUp - \ai_1 \GDown)}{\left(1+\sum_{i=2}^{\mathscr{M}} g_i\right)(\ai_1+\ei_1)\kappa}, & \GUp\geq\GThOne,\\
      0, & \text{otherwise}.
    \end{cases}   
\end{equation}

The solution \eqref{eq:dn} indicates a sharp transition above $\GThOne$ with a linear growth for increasing $\Delta \Gamma$, similarly as for the single mode discussed in Ref.~\cite{bennett2020symmetry}. The transition becomes less pronounced for a higher number of modes or larger degeneracies since $\Delta n$ is inversely proportional to $\mathscr{M}$ and $g_i$. Figure \ref{fig:Plot3} shows the dependence of $|\Delta n|$ for different values of $\Delta \Gamma$. Evidently, we see the breakdown of linearity of $|\Delta n|$ for larger values of $\Delta\Gamma$. If higher orders of $\kappa$ were retained, the correspondence between numerics and analytics would match at higher values of $\Delta\Gamma$. Note that our solution is only valid in the region $\GThOne\leq\GUp<\GThTwo$.

The solution \eqref{eq:n1_sol_dn_general} is similar to that in Ref.~\cite{bennett2020symmetry}, except for the correction factor in the denominator which depends on degeneracies and the number of modes. For an increasingly large system, the condensation transition becomes less distinct, similarly as for $\Delta n$. As a result, in the limit $\mathscr{M}\to \infty$ for a free photon gas, our findings indicate $\Delta n\to 0$ and $n_1 \to 0$ for a fixed $\GUp$. Since $\Delta n$ vanishes, the populations in any mode $i$ coincide with the ground mode, $n_i\to n_1$, up to a degeneracy factor.

In this Letter, we have demonstrated the emergence of a second condensation threshold in the regime of small mode spacing of a two dimensional photon gas in a planar cavity, which is inversely proportional to the mode spacing. The first and second thresholds are governed by system losses. At the second threshold, the ground mode becomes dominant in a different fashion compared to the case of larger mode spacing. From analytical and numerical solutions, we see how the phase transition disappears in the limit of a free 2D photon gas. We have found this to be caused by an infinite growth of the second threshold value. 

The intermediate region between the two thresholds, referred to as the supercooling region, is marked by incomplete condensation. Here, the separation in occupation between ground and first excited modes is linear in the mode spacing, hindering the system from reaching a fully condensed regime. Our analysis shows that potential uses of the photon BEC, such as the quantum sensor \cite{bennett2020symmetry}, will have to operate at a sufficiently large pumping rate to overcome the second threshold. Furthermore, our results are relevant for studies of open systems with a driven dissipative nature, as well as systems at the crossover towards continuum spectra.

We would like to thank Apurba Das, Dominik Lentrodt, Frieder Lindel, Axel U. J. Lode, Leon Espert Miranda, Axel Pelster, Andreas Redmann, Kirankumar Karkihalli Umesh, and Lucas Weitzel for fruitful discussions. This project has received funding from the European Union’s Horizon 2020 research and innovation programme under the Marie Skłodowska-Curie grant agreement number 847471. M.R. was supported by a DFG Research Grant No. 274978739.

\section*{SUPPLEMENTARY MATERIAL}
\setcounter{equation}{0}
\setcounter{figure}{0}
\renewcommand{\theequation}{S\arabic{equation}}
\renewcommand{\thefigure}{S\arabic{figure}}
In this Supplemental Material, we provide additional details of the paper, such as explicit expressions for the solutions and some additional figures, as well as motivate our method of finding the second threshold.

Here we briefly go through the method of finding both thresholds again and provide exact expressions for some solutions.
First, we show the solution of $n_2$ as a function of $n_1$.
We start with the rate equations for two modes:
\begin{align}\label{eq:adiabatic_rate_equation2}
\dot{n}_1=-\kappa n_1 +M \frac{\ei_1 \GTotUp (n_1+1)-\ai_1 \GTotDown n_1}{\GTotUp+\GTotDown},\\
\label{eq:adiabatic_rate_equation3}
\dot{n}_2=-\kappa n_2 +M \frac{\ei_2 \GTotUp (n_2+1)-\ai_2 \GTotDown n_2}{\GTotUp+\GTotDown}
\end{align}
with the total rates:
\begin{alignat}{2}
\GTotUp&=\GUp+ \ai_1 n_1+\ai_2 n_2,
\\
\GTotDown&=\GDown+ \ei_1 (n_1+1)+\ei_2 (n_2+1).
\end{alignat}
For a two-dimensional box potential, we have a nondegenerate ground mode. We also assume a nondegenerate excited state, $g_2=1$, since the final result will not depend on $g_2$. We now solve the steady state equation $\dot{n}_2=0$ and obtain:
\begin{align}\label{eq:n1twomodesol}
    &n_2(n_1)=-\frac{ A + B - \sqrt{C + (A + B)^2}}{2 \kappa(\GDownMTwo + \GUpMTwo)}
\end{align}
with the coefficients:
\begin{gather}
    A=M \GUpMTwo[\GDown+\GDownMOne (n_1+1)] - M\GDownMTwo (\GUp+n_1 \GUpMOne ),\\
    B=\kappa[\GDown +\GUp +\GDownMTwo  +\GDownMOne (n_1 +1) +n_1\GUpMOne] ,\\
    C=4 M \kappa \GDownMTwo (\GUp+n_1 \GUpMOne) (\GDownMTwo +\GUpMTwo ).
\end{gather}
The critical threshold is found by setting $\big(\partial^3 n_2 / \partial \GUp^3 \big )_{n_1}=0$ and solving for $\GUp$. The solution reads:
\begin{equation}\label{eq:GUpIntermediate}
    \GUp=\frac{F+ G \kappa -H \kappa ^2}{\left(\kappa - \ei_2 M\right)^2}
\end{equation}
with coefficients
\begin{gather}
    F=M^2  \ei_2 \left( \ai_2 \left[ \GDown+  \ei_1 (1+ n_1)\right]- \ai_1  \ei_2 n_1\right),\\
    \begin{split}
            G=M \Big[\big(2  \ai_1 n_1+ \GDown+ \ei_1 (1+ n_1)- \ei_2\big)  \ei_2 \\- \ai_2 \big( \GDown+ \ei_1 (1+ n_1)+2  \ei_2\big)\Big],
    \end{split}
\\
    H=\left( \GDown+ \ei_1+n_1 \left( \ai_1+ \ei_1\right)+ \ei_2\right).
\end{gather}
\begin{figure}
    \centering
    \includegraphics[width=0.4\textwidth]{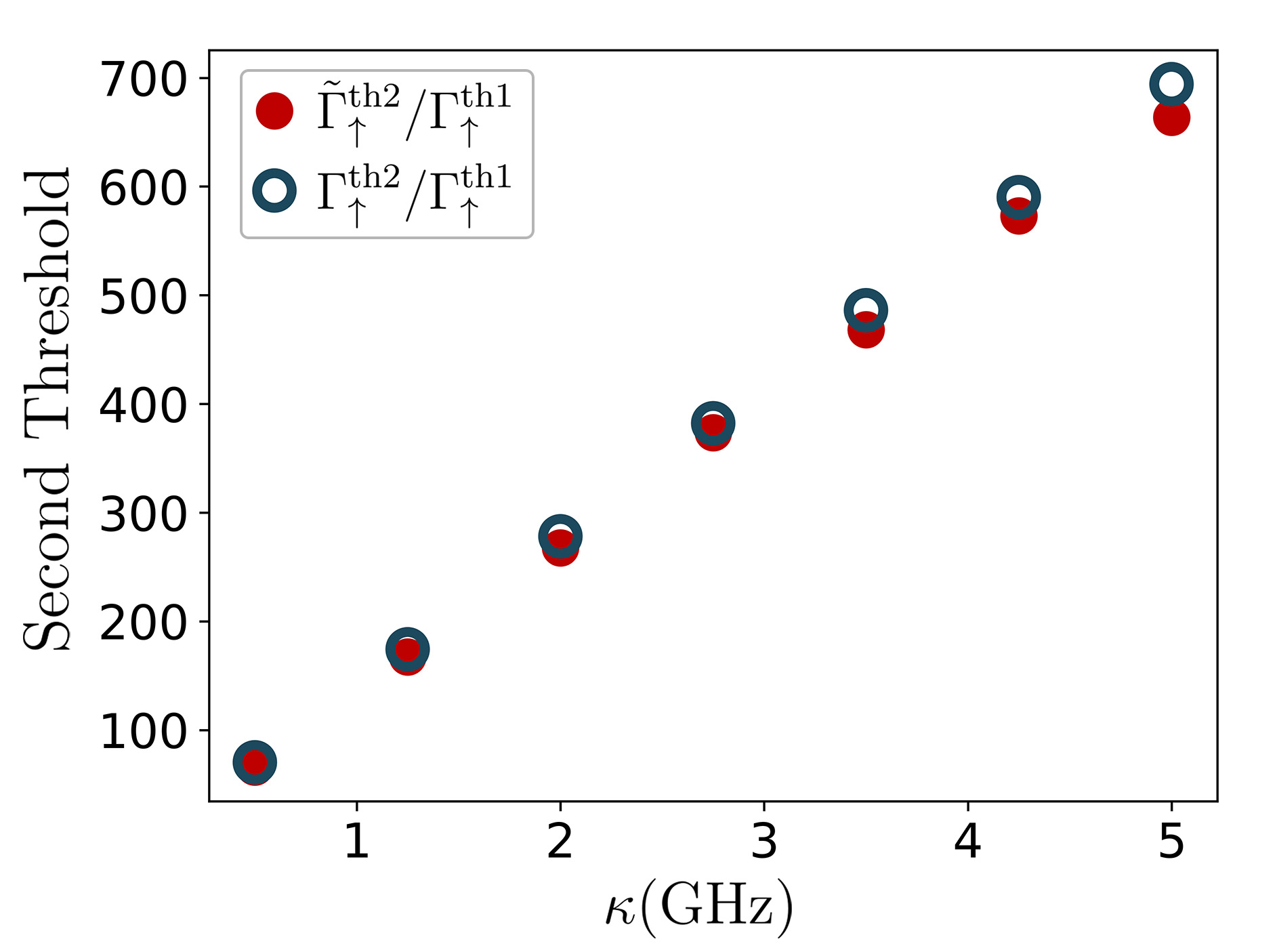}
    \caption{Both thresholds $\GThTwo/\GThOne$ (blue hollow dots) and $\GThTwoTilde/\GThOne$ (red filled dots) as a function of $\kappa$. Mode spacing is $\dkpar=50$ $\text{m}^{-1}$. }
    \label{fig:error_thresholds}
\end{figure}

Finally, in order to obtain a self-consistent expression of $\GUp$, we approximate $n_1$ as a solution from the single mode case. To obtain it, we solve the following equation:
\begin{align}
\dot{n}_1=-\kappa n_1 +M \frac{\ei_1 \GTotUp (n_1+1)-\ai_1 \GTotDown n_1}{\GTotUp+\GTotDown}
\end{align}
with the total rates:
\begin{alignat}{2}
\GTotUp&=\GUp+ \ai_1 n_1,
\\
\GTotDown&=\GDown+ \ei_1 (n_1+1).
\end{alignat}
The single-mode solution reads
\begin{equation}\label{eq:n1onemodesol}
    n_1=\frac{a-b + \sqrt{ c+(a-b)^2}}{2\kappa (\GDownMOne+\GUpMOne)}
\end{equation}
with
\begin{gather}
    a=M(\GDownMOne\GUp-\GDown\GUpMOne),\\
    b=\kappa(\GDown+\GDownMOne+\GUp),\\
    c=4\kappa M\GDownMOne\GUp(\GDownMOne+\GUpMOne).
\end{gather}
Both thresholds are then found by substituting Eq.~\eqref{eq:n1onemodesol} into Eq.~\eqref{eq:GUpIntermediate} and solving for $\GUp$. Utilizing the limit of small mode spacing, $\ai_2\approx \ai_1$ and $\ei_2\approx \ei_1$ and up to first order of $\kappa$ we obtain expressions of the first and the second threshold in the main paper.

The ground mode population $n_1$ in Eq.~\eqref{eq:n1onemodesol} for small $\kappa$ can be expanded in Taylor series and yields a piecewise solution, as in Ref.~\cite{bennett2020symmetry}:
\begin{equation}
n_1=
   \begin{cases}
      \displaystyle\frac{M(\ei_1 \GUp - \ai_1 \GDown)}{(\ai_1+\ei_1)\kappa}, & \GUp\geq\GThOne,\\
      0, & \text{otherwise}.
    \end{cases}   
\end{equation}

Taking the first derivative $\dd n_1 /\dd \GUp$ yields a step function, while the second derivative gives Dirac delta function at $\GUp=\GThOne$,  $\dd^2 n_1 /\dd \GUp^2 \sim \delta(\GUp-\GThOne)$. This observation is the motivation for our method of finding both thresholds for $\mathscr{M}\geq 2$.

In order to determine $\GThTwo$ we have evaluated the partial derivative $\big(\partial^3 n_2 / \partial \GUp^3\big)_{n_1}$, while for determining $\GThTwoTilde$ we have numerically evaluated the total derivative $\dd^3 n_2 / \dd \GUp^3$. We show the agreement between $\GThTwo$ and $\GThTwoTilde$ for $\mathscr{M}=2$ in the limit of small $\dkpar$, in Fig.~\ref{fig:error_thresholds} across the relevant range of $\kappa$ values.

\begin{figure}[h]
    \centering
    \includegraphics[width=0.4\textwidth]{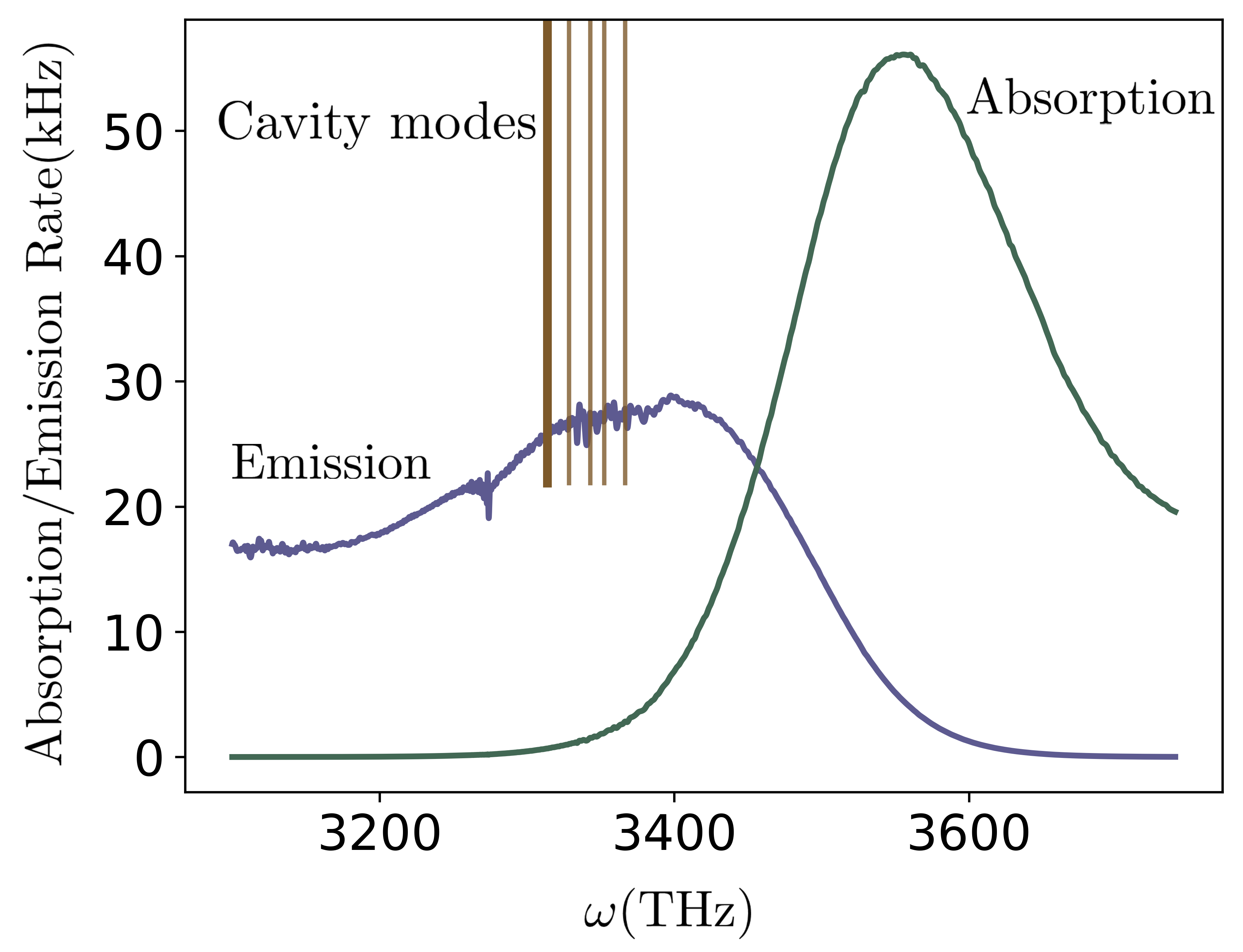}
    \caption{Absorption and emission spectra of Rhodamine 6G molecule. Vertical lines indicate cavity modes (mode spacing is scaled up for visualization purposes). The thick vertical line indicates the ground mode.}
    \label{fig:spectrum_paper}
\end{figure}

Figure \ref{fig:spectrum_paper} shows the absorption and emission spectra as a function of angular frequency \cite{julian2024private} together with the occupied cavity modes.
The absorption curve was obtained from measurements and the emission curve was obtained from the absorption using the Kennard--Stepanov relation $\ei_i=\ai_i e^{-\beta \hbar (\omega_i-\omega_{\text{ZPL}})}$.


\providecommand{\noopsort}[1]{}\providecommand{\singleletter}[1]{#1}%

\end{document}